\documentclass[conference]{IEEEtran}
\usepackage{cite}
\usepackage{graphicx}
\usepackage[cmex10]{amsmath}
\usepackage{array}
\usepackage{fixltx2e}
\usepackage{algorithmic}
\usepackage{algorithm}
\usepackage{amsfonts}
\usepackage{amssymb}

\newtheorem{lemma}{Lemma}
\newtheorem{defi}{Definition}
\newtheorem{exam}{Example}
\newtheorem{theo}{Theorem}

\newcommand{\ou}{\ensuremath{\overline{u}}}
\newcommand{\ox}{\ensuremath{\overline{x}}}
\newcommand{\ot}{\ensuremath{\overline{t}}}
\newcommand{\oy}{\ensuremath{\overline{y}}}
\newcommand{\oz}{\ensuremath{\overline{z}}}

\newcommand{\rulefrac}[2]{
\frac{\rule[-3pt]{0pt}{10pt}\mbox{$\ #1\ $}}{\rule{0pt}{10pt}\mbox{$\ #2\ $}}
}

\begin{document}

\title{Cyclic and Inductive Calculi are Equivalent}

\author{\IEEEauthorblockN{R\u azvan Voicu and Mengran Li}
\IEEEauthorblockA{School of Computing\\
National University of Singapore\\
Email: \{razvan,limengra\}@comp.nus.edu.sg}
}
\maketitle

\begin{abstract}
Brotherston and Simpson [citation] have
formalized and investigated cyclic
reasoning, reaching the important conclusion
that it is at least as powerful
as inductive reasoning (specifically, they showed that each
inductive proof can be translated into a
cyclic proof). We add to their investigation
by proving the converse of this result, namely
that each inductive proof can be translated into an
inductive one. This, in effect, establishes the
equivalence between first order cyclic and inductive
calculi.
\end{abstract}

\section{Introduction}

Cyclic proofs are often a convenient replacement for induction.
They do not require the discovery of induction hypotheses that are
usually stronger than the conjecture at hand, but rather proceed with
simple case analysis steps and applications of simple laws,
allowing ``smaller'' instances of the original conjecture to be used
as lemmas. While cyclic proof methods still require a high level of
ingenuity, they appear to lead to clearer proofs,
and seem to lead to more successful automation when implemented
in automated reasoning tools.

Under these circumstances, formal investigation of cyclic reasoning
becomes very important. In \cite{BrotherstonJLC}, Brotherston and
Simpson have provided a rigorous and sound treatment of first-order
infinite
descent reasoning, identifying cyclic proofs as a special case of
practical importance. The
infinite descent calculus
in \cite{BrotherstonJLC}
 is shown to be sound and complete; 
most importantly, cyclic proofs are shown to be
a valid replacement for inductive proofs, via a
translation scheme that converts 
every inductive proof
into a cyclic one. While this would be sufficient to
justify the use of cyclic reasoning in practice, 
it does not complete its formal investigation.

In the present paper, we prove Conjecture~7.7
of \cite{BrotherstonJLC}, which states that
every cyclic proof can be converted into
an inductive one. The proof is subtle and laborious, and while of
possibly no immediate practical importance, it establishes the
equivalence of inductive and cyclic reasoning,
hopefully making a contribution towards a more wide-spread
adoption of the latter in mechanical and
automated reasoning tools. 


Our paper is organized as follows.
Section 2 discusses related work, giving a brief
overview of the developments in \cite{BrotherstonJLC}.
Section 3 introduces the notations and terminology that will be used
in the rest of the paper. Section 4 presents the proof
that each cyclic proof can be converted into an
inductive one. 
Section 5 concludes.

\section{Related Work}

The present development builds heavily on Brotherston and Simpson's work
\cite{conf/tableaux/brotherston05,
thesis/brotherston,
brotherston07lics,
BrotherstonJLC}, which we 
describe it here in some detail.
The work is structured in four main parts. First,
a first order language is defined, in the spirit of
\cite{Barwise77}, and then augmented with inductive
definitions based on Martin-L\"of's ``ordinary productions''
\cite{conf/sls/martinlof71}. A standard interpretation is defined
for this language, where the inductive component is interpreted as
the least fixed point of a monotone operator,
cf. \cite{bookchapter/handbook/Aczel}\footnote{
A Henkin semantics is also defined at this point, and used to prove a
completeness result for the infinite descent calculus; this, however,
falls outside the scope of this paper.}.
Second, an inductive sequent calculus, named LKID, is defined.
This calculus adopts the original rules of Gentzen's LK calculus
\cite{book/gentzen}, and adapts its induction rules from
\cite{conf/sls/martinlof71}. The calculus is proved to be sound and
complete with respect to Henkin semantics, 
and to enjoy a cut-elimination property, due to the
particular formulation of the induction rules, which allow new
formulas to be introduced in a proof as induction hypotheses. Due to
this peculiarity, LKID no longer enjoys the {\it subformula property}
that is instrumental in proving several significant properties of
the LK calculus.
Third, an infinite descent calculus, LKID$^\omega$, is
introduced. This
calculus is obtained by weakening the induction rules 
of LKID into {\it case-split} rules, which essentially no longer have
a placeholder for an induction hypothesis. However, proofs are now
allowed to be infinite, so long as every infinite branch in the proof
tree exhibits a {\it progress condition}, namely that case-split rules
of inductive predicates from the same set of mutually inductive
predicates occur infinitely often along the path.
LKID$^\omega$ is proved sound and complete and, 
due to its potentially infinite proofs, stronger than LKID 
(but also interesting only from a theoretical perspective).
Fourth, a finitary restriction is identified for LKID$^\omega$ proofs,
where proof trees are regular, and can thus be represented as graphs with
possible cycles; each infinite path will yield a cycle in the graph
representation, and the progress condition translates into every cycle
going accross at least one application of a case-split rule. 
The restricted calculus, of practical importance due to its finitary
nature, is denoted as CLKID$^\omega$;
its proofs are called {\it cyclic proofs}, and represent the focus of
the current paper. The cyclic calculus is proved to be sound, and at
least as powerful as the inductive calculus LKID; the proof is
achieved via a translation scheme that converts every inductive proof
into a cyclic one. Conjecture 7.7 in
\cite{BrotherstonJLC} hypothesizes that the converse is also true,
that is, every  CLKID$^\omega$ proof can be converted into a LKID
proof. Our paper presents a proof of this conjecture.

An approach similar to cyclic reasoning
is presented for the $\mu$-calculus in
\cite{conf/FOSSACS/Sprenger03,journals/TIA/Sprenger03}.
A comprehensive overview of infinite descent, and approaches to
using it in proof automation are given in
\cite{conf/birthday/Wirth05,journals/corr/abs-0902-3623}.
Cyclic reasoning-like methods have also been used in automated
provers, especially for program verification, cf. 
\cite{Nguyen07,BartheS-RTA03,
conf/tableaux/Stratulat05,conf/cp/JaffarSV08}.

\section{Preliminaries}

\subsection{First-order logic with inductive definitions 
}
Following \cite{BrotherstonJLC},
we consider a fixed language $\Sigma$ with inductive predicate symbols
$P_1,\ldots,P_n$. Terms of $\Sigma$ are defined as usual; we write
$t(x_1,\ldots,x_k)$ for a term whose variables are contained
in $\{x_1,\ldots,x_k\}$. The formulas of the logic
are the usual formulas of first-order logic\footnote{
We do not define equality as part of our language; an equality
predicate may be defined inductively if needed.
}.
We denote by $\overline{x}$ the sequence of variables
$x_1,\ldots,x_k$, and by $\overline{t}(\overline{x})$
the sequence of terms $t_1(x_1,\ldots,x_k),\ldots,
t_l(x_1,\ldots,x_k)$; the values $k$ and $l$
will be usually understood from the context.
\bigskip

\begin{defi}[Inductive definition set]
\label{inductivedef}
An inductive definition set $\Phi$ for
$\Sigma$ is a finite set of productions
of the form:
$$
\frac{
Q_1(\ou_1(\ox))\ldots
Q_k(\ou_k(\ox))\ 
P_{j_1}(\ot_1(\ox))\ldots
P_{j_m}(\ot_m(\ox))}{
P_i(\ot(\ox))}\quad(1)
$$
where $Q_1,\ldots,Q_k$ are ordinary predicate symbols,
$j_1,\ldots,j_m,i\in\{1,\ldots,n\}$, 
$\ox$ is a set of distinguished variables, and 
the lengths of the sequences
$\ou_1(\ox),\ldots,\ou_k(\ox),\ot_1(\ox),\ldots,\ot_m(\ox),\ot(\ox)$ 
appropriately match the arities of the predicate
symbols.
\end{defi}
\bigskip

The following example introduces inductive predicates
that shall be later used in our example proofs. In our
examples, we prefer a functional notation for our predicates,
that is, we write $(P\, t_1\,\ldots\,t_k)$ instead of
$P(t_1,\ldots,t_k)$.
\bigskip

\begin{exam}
We define the predicates {\it nat} and {\it tplus},
representing natural numbers, and tail-recursive addition.
$$
\frac{}{{\it nat}\ 0}\quad
\frac{{\it nat}\ x}{{\it nat}\ (S\,x)}\quad
\frac{{\it nat}\ x}{{\it tplus}\ 0\,x\,x}\quad
\frac{{\it tplus}\ x\, (S\, y)\, z}{{\it tplus}\ (S\,x)\,y\,z}
$$
\end{exam}

\subsection{A proof system for induction (LKID)}

We write sequents of the form
$\Gamma\vdash\Delta$,
where $\Gamma$ and $\Delta$ are
finite sets of formulas, and use
the notation $\Gamma[\theta]$
to mean that the substitution $\theta$
of terms for free variables is applied
to all formulas in $\Gamma$.
For first order logic,
we use the standard sequent calculus rules
given in Figure~\ref{lkidrules}.
However, we do not add equality rules to our incarnation
of this calculus, since an equality predicate,
can be defined via inductive rules, if needed.
We read the comma in sequents as set union,
and thus the contraction and exchange rules are
not necessary.

\begin{figure*}
\fbox{
\begin{minipage}{0.96\textwidth}
\bigskip

\input{lkidrules}
\caption{LKID rules}\label{lkidrules}
\end{minipage}}
\end{figure*}

We augment the rules in  Figure~\ref{lkidrules}
with rules for introducing atomic formulas
involving inductive predicates on the left and
right of sequents.

First, for each production
$\Phi_{i,r}\in \Phi$,
each necessarily in the format of
Definition~\ref{inductivedef},
we include a corresponding sequent calculus
right introduction rule for $P_i$:
$$
\frac{
\begin{array}{l}
\Gamma\vdash Q_1(\ou_1(\oy)),\Delta\ \ldots\ 
\Gamma\vdash Q_k(\ou_k(\oy)),\Delta\\ 
\hspace*{.5in}\Gamma\vdash P_{j_1}(\ot_1(\oy)),\Delta\ \ldots\ 
\Gamma\vdash P_{j_m}(\ot_m(\oy)),\Delta
\end{array}}{
\Gamma\vdash P_i(\ot(\oy)),\Delta}\,(P_iR_r)
$$
Here $\oy$ is assumed to be a sequence of terms of the
same length as the sequence $\ox$ of variables 
explicitly identified as
occurring in the production, and the occurrences of \oy in the rule
above represent the substitution $[\oy/\ox]$.

The left-introduction rules for inductively defined predicates
represent in fact induction rules. The definition of these rules
requires the notion of {\it mutual dependency} as defined in
\cite{conf/sls/martinlof71}, 
and reproduced in the following definition:
\bigskip

\begin{defi}[Mutual dependency]
Let ${\cal R}\subseteq\{P_1,\ldots,P_n\}$ $\times\{P_1,\ldots,P_n\}$
be defined as the least relation satisfying the property
that $(P_i,P_j)\in {\cal R}$
whenever there exists a rule in $\Phi$ such that
$P_i$ occurs in the rule's conclusion,
and $P_j$ occurs among its premises.
Denote by ${\cal R}^*$ the reflexive-transitive closure of $\cal R$.
Then, two predicates $P_i$ and $P_j$ are {\em mutually
dependent} if both $(P_i,P_j)\in {\cal R}^*$
and $(P_j,P_i)\in {\cal R}^*$.
\end{defi}
\bigskip

To obtain an instance of the left-introduction rule
for any inductive predicate $P_j$, we first associate
with every inductive predicate $P_i$ a tuple
$\oz_i$ of size equal to the arity of $P_i$.
Furthermore, we associate to every predicate
$P_i$ that is mutually dependent with $P_j$ an
{\em induction hypothesis} $F_i$ (which is, essentially,
an arbitrary formula), possibly containing some of the
induction variables $\oz_i$. Then, we define
for each $i\in\{1,\ldots,n\}$:
$$
G_i = \left\{\begin{array}{ll}
F_i & \mbox{if }P_i\mbox{ and }P_j
\mbox{ are mutually dependent}\\
P_i(\oz_i) & \mbox{otherwise}
\end{array}\right.
$$
An instance of the induction rule for $P_j$
has the following schema:
$$
\frac{\mbox{minor premises}\quad
\Gamma,F_j[\ou/\oz_i]\vdash\Delta}{
\Gamma,P_j(\ou)\vdash\Delta}\,
(P_j{\it IND})
$$
where the premise $\Gamma,F_j[\ou/\oz_i]\vdash\Delta$ is called
the {\em major premise} of the rule instance,
and for each production
of $\Phi$, of the form (1), having in its conclusion a predicate
$P_i$ that is mutually dependent with $P_j$, 
there is a corresponding minor premise:
$$
\begin{array}{l}
\Gamma,Q_1(\ou_1(\oy)),\ldots,Q_k(\ou_k(\oy)),\\
\hspace*{0.3in}
G_{j_1}[\ot_1(\oy)/z_{j_1}],\ldots,G_{j_m}[\ot_m(\oy)/z_{j_m}]\\
\hspace*{1in}\vdash F_i[\ot(\oy)/z_i],\Delta
\end{array}
$$
where $\oy$ is a vector of distinct fresh variables (i.e. not
occurring in 
$FV(\Gamma\cup\Delta\cup\{P_{j_l}(\ot_l(\ox))|l=1,\ldots,m\})$)
of the same
length as the vector $\ox$ of variables explicitly
identified in the production.

All the formulas of the calculus LKID, with the exception of
the cut rule, have a {\it principal formula}, which is, by
definition, the formula occurring in the lower sequent of the
inference which is not in the cedents $\Gamma$ and $\Delta$. Every
inference, except weakenings, has one or more {\it auxilliary formulas},
which are the formulas $A$ and $B$, occurring in the upper
sequent(s) of the inference. The formulas which occur in the cedents
$\Gamma$ or $\Delta$ are called {\it side formulas}
of the inference. The two auxilliary formulas of a cut inference are
called {\it cut formulas}.
\bigskip

\begin{exam}[Induction rule for natural numbers]
$$
\frac{
\begin{array}{l}
\Gamma \vdash\\
 F(0),\Delta
\end{array}\quad
\begin{array}{l}
\Gamma, F(p) 
\vdash \\
F(S(p)),\Delta
\end{array}\quad
\begin{array}{l}
\Gamma,F(t) \\
\vdash \Delta
\end{array}
}{\Gamma,nat\,t\vdash \Delta
\makebox[0in][l]{{}\hspace*{0.9in}\it (natIND)}
}
$$
where $F(z)$ is the induction hypothesis, 
parameterized on an induction variable
$z$, and $p$ is a fresh variable, i.e.
$p\not\in FV(\Gamma\cup\Delta\cup F(z))$.
\end{exam}
\bigskip

\begin{exam}[Induction rule for {\it tplus}]
$$
\frac{
\begin{array}[b]{l}
\Gamma,\\
nat\,u\vdash\\
F(0,u,u),\Delta
\end{array}\ \ 
\begin{array}[b]{l}
\Gamma,\\
F(u,S\, v,w)\vdash\\
F(S\, u,v,w), \Delta
\end{array}\ \ 
\begin{array}[b]{l}
\Gamma, \\
F(t_1,t_2,t_3)\\
\vdash \Delta
\end{array}
}{
\Gamma, {\it tplus}\,t_1\,t_2\,t_3
\vdash\Delta
\makebox[0in][l]{{}\hspace*{0.6in}\it (tplusIND)}
}
$$
where $F(z_1,z_2,z_3)$ is the induction hypothesis, 
parameterized on induction variables
$z_1,z_2,z_3$, and $u,v,w$ are fresh variables, i.e.
$u,v,w\not\in FV(\Gamma\cup\Delta\cup F(z_1,z_2,z_3))$.
\end{exam}
\bigskip

A {\em derivation tree}
is a tree of sequents in which each parent is obtained as the
conclusion of an inference rule with its children as premises.
A {\it proof} in LKID is a finite derivation tree all of 
whose branches end in an axiom (i.e. a proof rule without premises).
\bigskip

\begin{figure*}
\frame{
\begin{minipage}{0.96\textwidth}
\bigskip

\begin{center}
\includegraphics[scale=0.75]{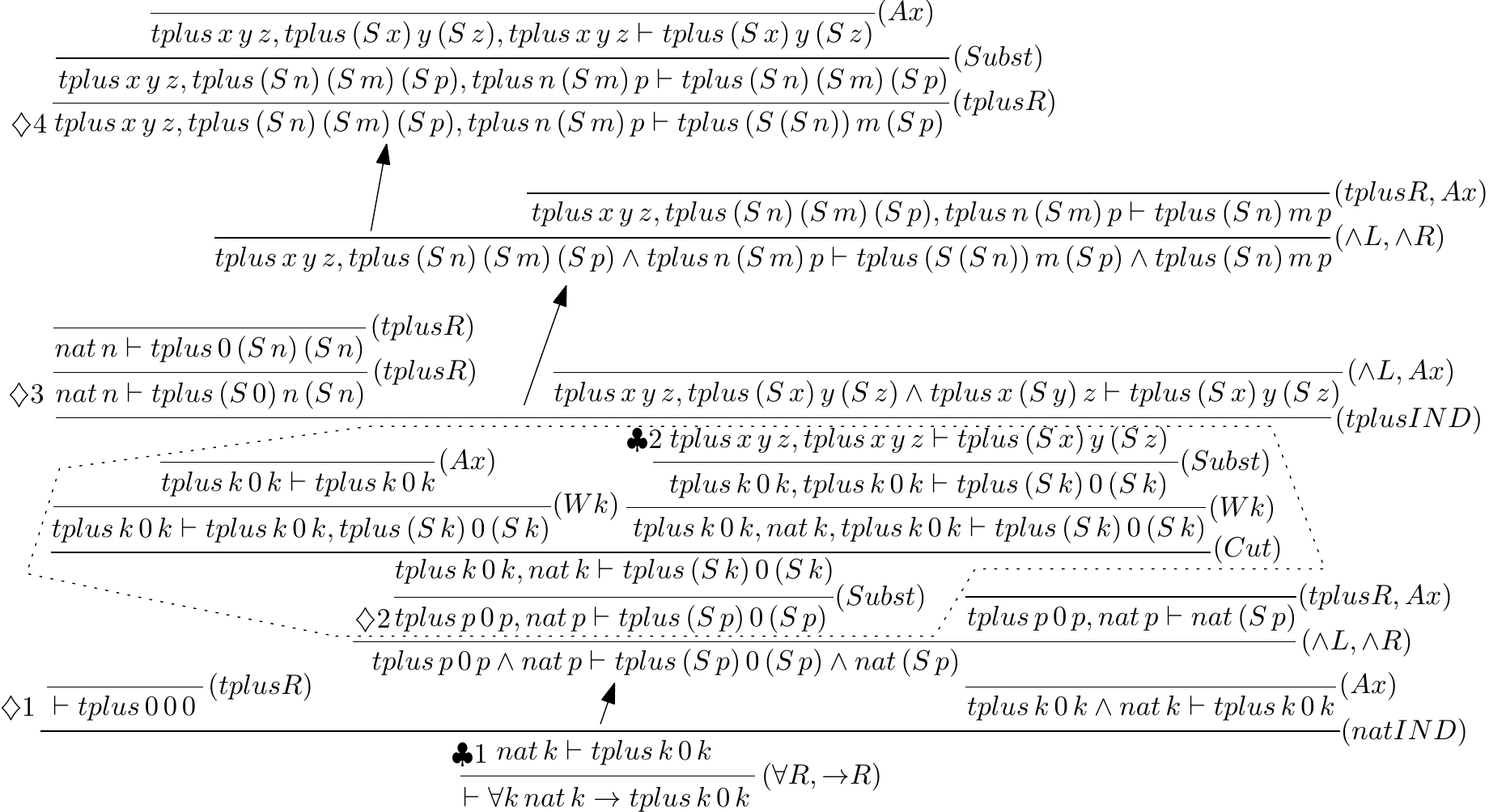}
\end{center}

\caption{Inductive Proof}\label{tplusinductiveproof}
\end{minipage}

}
\end{figure*}

\begin{exam}
Figure~\ref{tplusinductiveproof}
depicts the proof tree for the sequent
$
nat\,k\vdash {\it tplus}\ k\,0\,k
$. The induction is nested.
To obtain the proof, we first apply
the (natIND) rule, with the 
induction hypothesis ${\it tplus}\,z\,0\,z$.
The first and third premise can then be discharged
immediately, whereas the second premise
requires further induction on the predicate
{\it tplus}. For the second round of induction,
the induction hypothesis is
${\it tplus}\,z_1\,(S\,z_2)\,z_3$.
\end{exam}

\subsection{A cyclic proof system CLKID$^\omega$}

The proof rules of CLKID$^\omega$ are the rules of LKID, except that
for each inductive predicate $P_i$ of $\Sigma$, the induction rule
$(P_i{\it IND})$ is replaced by the {\em case-split rule}
$$
\frac{\mbox{case distinctions}}{
\Gamma[\ou/\oy],P_i(\ou)\vdash\Delta[\ou/\oy]}\ (P_iL)
$$
where for each production of the form $(1)$ having predicate
$P_i$ in its conclusion, there is a corresponding
case distinction
$$
\begin{array}{l}
\Gamma[\ot(\ox)/\oy],Q_1(\ou_1(\ox)),\ldots,Q_k(\ou_k(\ox)),\\
\hspace*{0.5in}P_{j_1}(\ot_1(\ox)),\ldots,P_{j_m}(\ot_m(\ox))
\vdash\Delta[\ot(\ox)/\oy]
\end{array}
$$
where $\oy$ is a vector of distinct fresh
variables of the same length as $\ox$.
\bigskip

\begin{exam}[Case-split for nat and tplus]
The rules {\it (natL)} and {\it (tplusL)}
represent the case-split rules for the
predicates {\it nat} and {\it tplus}.

$$
\frac{
\Gamma[0/y]\vdash\Delta[0/y]
\quad
\Gamma[S\,p/y],nat\,p \vdash\Delta[S\,p/y]
}{
\Gamma[t/y],nat\, t\vdash\Delta[t/y]
}\quad{\it (natL)}
$$

$$
\frac{
\begin{array}{c}
\Gamma[0/y_1,x/y_2,x/y_3]\\
\vdash\\
\Delta[0/y_1,p/y_2,p/y_3]
\end{array}
\quad
\begin{array}{c}
\Gamma[S\,x/y_1,y/y_2,z/y_3],\\
{\it tplus}\,x\,(S\,y)\,z
\vdash\\
\Delta[S\,x/y_1,y/y_2,z/y_3]
\end{array}\quad
{\it (tplusL)}\\
}{
\Gamma[t_1/y_1,t_2/y_2,t_3/y_3],{\it tplus}\,t_1\,t_2\,t_3
\vdash \Delta[t_1/y_1,t_2/y_2,t_3/y_3]
}
$$
\end{exam}
\bigskip

The notion of proof for the CLKID$^\omega$
depends on a progress condition that needs
to distinguish between multiple occurences of
the same inductive predicate in a sequent.
To make that possible, we tag induction predicates
by natural numbers; thus $P^\alpha$ 
denotes the {\it tagged} version of inductive predicate
$P$, where $\alpha$ is a natural number. 
A sequent where all inductive predicate occurrences
are tagged is called a {\it tagged sequent}.
\bigskip

\begin{defi}[Tagged proof tree]
A {\it tagged proof tree} is a proof tree where each sequent is tagged
according to the following rules:
\begin{itemize}
\item The endsequent is tagged such that each inductive
predicate occurrence has a distinct tag.
\item Inferences preserve the tags of predicates in side formulas.
\item If an inductive predicate $P^\alpha$ is the principal formula in
  an inference, the predicates in the auxiliary formulas inherit the
tag $\alpha$.
\item In a cut inference, the cut formula may be a tagged predicate that
  appeared as a formula in the antecedent of an ancestor node. 
\item General
  formulas may appear in cut inferences only if their inductive
  predicates have fresh tags (i.e. not appearing at ancestor nodes).
\end{itemize}
\end{defi}
\bigskip

\begin{defi}[Proof in CLKID$^\omega$]
A proof in CLKID$^\omega$ is a tagged derivation tree
where each leaf is either (a) the tagged version of an axiom; or 
(b) it is {\it cyclic}, that is, 
it contains a sequent that is identical (tags included)
to one of its ancestors
called {\em companion node}, and with the added
property that on the path from a companion to its corresponding
cyclic leaf
there is at least one application of a case-split rule. 
\end{defi}
\bigskip

Our notion of cyclic proof is slightly different, but equivalent to
the notion of {\it normalized cyclic proof} 
in \cite{BrotherstonJLC}. In the progress condition, the use
of tags ensure that multiple occurrences of the same inductive
predicates are distinguished, so that a cycle is obtained by repeated
case analysis of inductive predicates from the same set of mutually
inductive predicates. In \cite{BrotherstonJLC}, this is ensured by
defining the notion of {\it trace} for infinite LKID$^\omega$ proofs.
The progress condition can be relaxed in a variety of ways, leading to
simpler proofs in practice; however,
this is outside the scope of this work.

In what follows, we shall assume that proof trees are implicitly
tagged, and we shall not mention the tags unless they are relevant to
the context.

We denote proofs by ${\cal D}=({\cal S},{\cal R},{\cal P},{\cal C})$,
where $\cal S$ is the set of sequents appearing in $\cal D$,
$\cal R$ is function mapping every sequent 
$\Gamma\vdash\Delta\in\cal S$ to the inference rule
that was applied to it in the proof,
$\cal P$ is a function mapping every sequent to its set of premises,
and $\cal R$ is a partial function mapping cyclic leafs into their
companions.\bigskip

\begin{figure*}
\frame{
\begin{minipage}{0.96\textwidth}
\bigskip

\begin{center}
\includegraphics[scale=0.8]{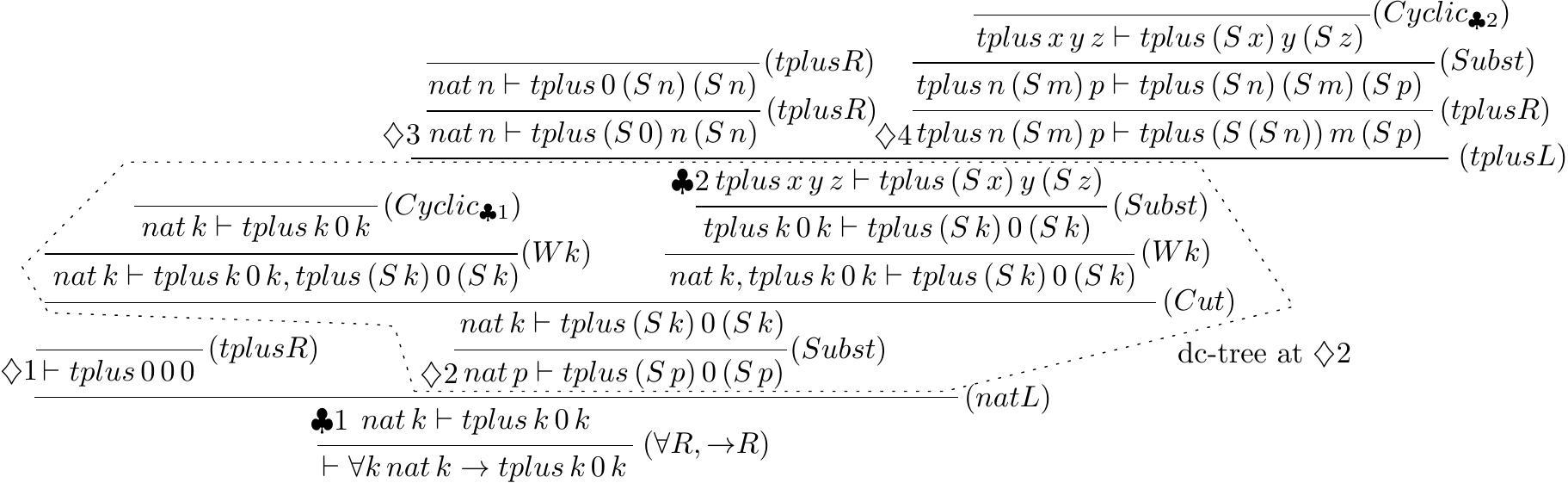}
\end{center}

\caption{Cyclic Proof}\label{tpluscyclicproof}
\end{minipage}

}
\end{figure*}

\begin{exam}[Cyclic proof]
Figure~\ref{tpluscyclicproof}
depicts a cyclic proof tree for the sequent
$\forall k\ nat\,k \rightarrow {\it tplus}\,k\,0\,k$.
In this tree, the leaf nodes where the notations
$({\it Cyclic}_{\clubsuit1})$ and $({\it Cyclic}_{\clubsuit2})$
appear as rule names represent cyclic leafs, and
the nodes marked with $\clubsuit1$ and $\clubsuit2$
represent the corresponding companion nodes. With this example, we want
to explore how we can obtain an inductive proof from a cyclic one. Intuitively,
as we start from the end sequent, we can mimic the application of rules in the
cyclic proof to build the inductive proof, up to the point where we encounter
a casesplit rule. The only choice there is to apply an induction rule. A possible
candidate for the induction hypothesis is $tplus \, p \, 0 \, p$. But that's not
good enough because we want to have a correspondance betwen the vertices in the
inductive proof and the vertices in the cyclic proof, where for each vertex $v$
in the inductive proof that is not a translation artefact, there is a $v'$ in the
cyclic proof such that if the sequent at $v$ is $\Gamma \vdash \Delta$, then
the sequent at $v'$ is $\Gamma' \subseteq \Gamma, \vdash \Delta$. And clearly
$tplus \, p \, 0 \, p$ alone will not give us $nat \, p$ at $\diamondsuit 2$.
Therefore we set the induction hypothesis to $tplus \, p \, 0 \, p \wedge nat \, p$.
This generates a translation artefect, $tplus \, p \, 0 \, p, nat \, p \vdash nat \, (S \, p)$,
which can be easily discharged by a right introduction of $nat$ followed by Axiom.
We continue building the inductive proof by mimicing the cyclic one. For example, for
the premise $\diamondsuit2$ in the inductive proof, we can get a derivation tree marked
by the dotted polygon, that is isomorphic to the one marked by the dotted polygon in the
cyclic proof. There are two leaf nodes in this dotted tree. One corresponds to the
node where the cyclic rule is applied, which we can discharge by Axiom. The other corresponds
to another casesplit in the cyclic proof. We apply induction in the same way and complete
the proof.

\end{exam}



\section{Equivalence of LKID and CLKID$^\omega$}

In this section we present the main result of our paper,
consisting of a translation scheme that converts
cyclic proofs into inductive ones, and prove its correctness.
We first introduce the necessary terminology.

Given a cyclic proof tree ${\cal D}=({\cal S},{\cal R},{\cal P},{\cal
  C})$, we say that a node $\Gamma\vdash\Delta$ is a {\it case-split
  node}, or a {\it c-node} if ${\cal R}(\Gamma\vdash\Delta)$
is a case-split rule with the inductive predicate $P_i$
as a principal formula; we shall denote
$P_i$ as ${\it cs}(\Gamma\vdash\Delta)$. We call the premises
of a {\it c}-node {\it descendants of case-split nodes}, or {\it d-nodes}.
By abuse of notation, we shall use ${\it cs}(\Gamma\vdash\Delta)$
to denote the principal formula of the inference that has
$\Gamma\vdash\Delta$ as a premise when
$\Gamma\vdash\Delta$ is a $d$-node.
In our examples we shall mark
$c$- and $d$-nodes by the symbols
$\clubsuit2$ and $\diamondsuit2$, respectively.
The {\it {\bfseries d}escendant-of-case-split-rooted-up-to-new-%
{\bfseries c}ase-split tree}, 
or {\it dc-tree}, rooted at a {\it d}-node $\Gamma_0\vdash\Delta_0$ is the
derivation tree obtained from the subtree of $\cal D$ whose root
is $\Gamma_0\vdash\Delta_0$, by removing all subtrees rooted at
all {\it d}-nodes distinct from $\Gamma_0\vdash\Delta_0$.
That is, no internal node of a dc-tree is a {\it c}-node, and its frontier
contains only leafs of $\cal D$ and {\it c}-nodes. 
Given a {\it dc}-tree $\cal T$, 
we
say that {\it $\cal T$'s rules are applicable} 
(or, for short, that {\it $\cal T$
is applicable}) to a sequent $\Gamma\vdash\Delta$
if a correct derivation tree ${\cal T}'$ 
can be formed with the endsequent
$\Gamma\vdash\Delta$, such that there exists
a isomorphism between $\cal T$ and $\cal T'$
that preserves the rules and their principal formulas 
accross every pair of isomorphic elements.

If a node is both a $d$- and a $c$-node, then the $dc$-tree
rooted at this node is empty. An empty $dc$-node is applicable to
any sequent, leaving it unchanged.
\bigskip

\begin{exam}[dc-tree]
In Figure~\ref{tpluscyclicproof}, the dotted polygon encircles
a $dc$-tree rooted at the sequent marked by $\diamondsuit2$.
On its frontier, there is a $c$-node, marked with
$\clubsuit2$. In Figure~\ref{tplusinductiveproof},
the dotted polygon represents the result of applying the $dc$-tree
given in the previous figure to the sequent
${\it tplus}\,p\,0\,p,\,{\it nat}\,p\vdash
{\it tplus}\,(S\,p)\,0\,(S\,p)$. We note that the two encircled
derivation trees have the same shape, and the same rules
at isomorphic nodes.
\end{exam}
\bigskip

We denote by ${\cal F}(P(\ot))$ the set
of predicates that are mutually
inductive with $P(\ot)$. If $P(\ot)$
is tagged, the predicates in 
${\cal F}(P(\ot))$ will have the same tags.
The set of inductive predicates occurring in the
antecedents of sequents of a proof tree may
be partitioned into several disjoint families of mutually 
inductive, identically tagged predicates.
We shall denote the set of such families by
${\it Fam}({\cal D})$, for a given cyclic proof $\cal D$.
Given a sequent $\Theta\equiv\Gamma,P(\ot)\vdash\Delta$,
we call ${\cal E}(\Theta,P(\ot))\equiv
\bigwedge\Gamma\rightarrow\bigvee\Delta$
the {\it extract} of $\Theta$ w.r.t. $P(\ot)$.
The newly introduced conjunction, disjunction, and
implication connectives are called {\it distinguished
connectives}, and shall have special treatment in
our translation scheme, as compared to the original
connectives that appear in $\Gamma$ and $\Delta$.

A cyclic proof is in {\it canonical form} if all its cyclic leafs are
{\it c}-nodes. A non-canonical proof can be easily converted by first
unwinding the proof tree by pasting a
copy of each companion-rooted subtree to the coresponding cyclic
leaf. Doing this {\it ad infinitum} will result in an infinite tree,
representing in fact an LKID$^\omega$ proof tree,
cf. \cite{BrotherstonJLC}. On each infinite path, some {\it c}-node will occur
infinitely often. Now, cut a finite ``tip'' of the infinite tree, so
that its frontier cuts accross the second occurrence
of the infinitely-often occurring {\it c}-node on every infinite 
path. Then, the finite ``tip'' is a cyclic proof in
canonical form.

\begin{figure*}
\frame{
\begin{minipage}{0.96\textwidth}
\bigskip

\begin{center}
\includegraphics[scale=0.75]{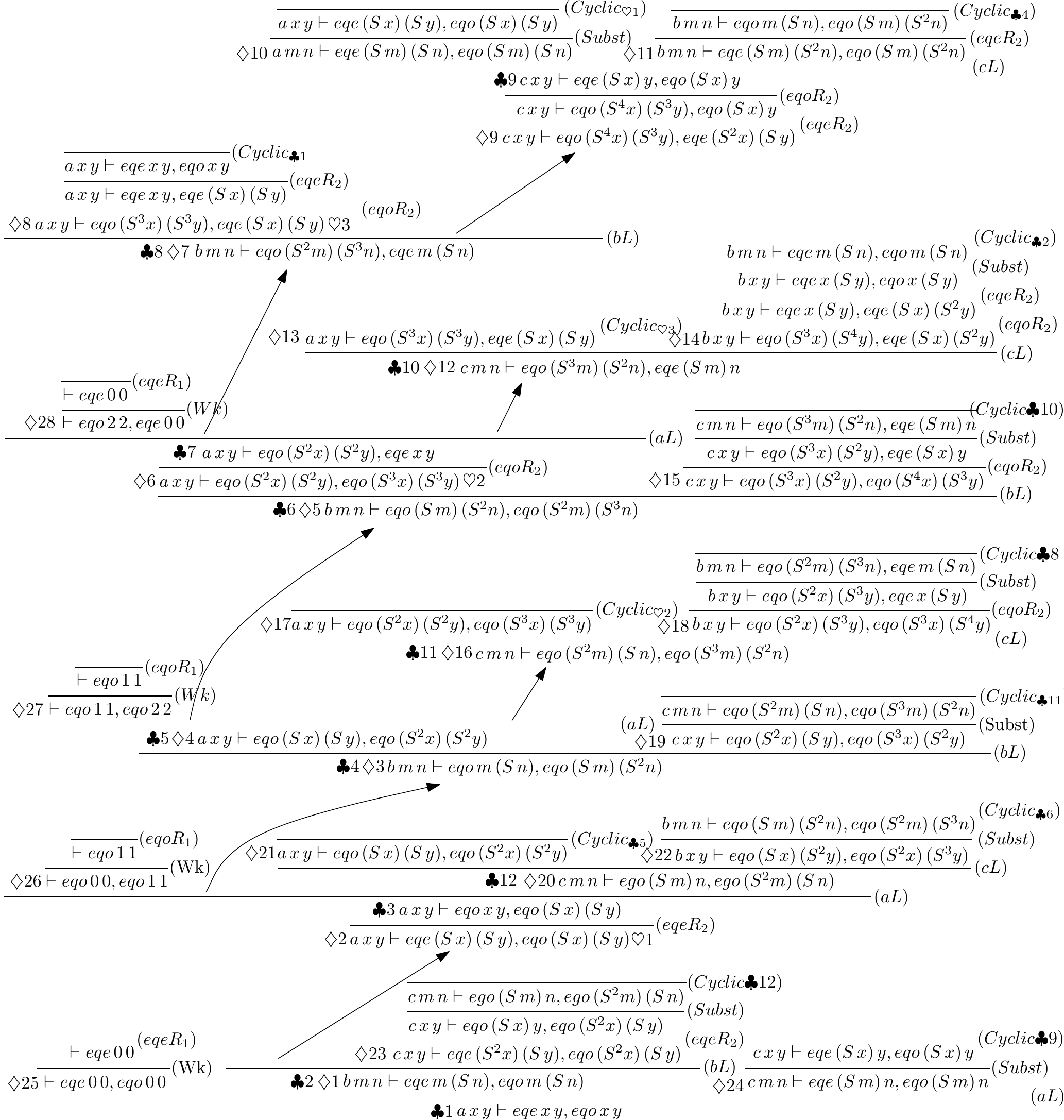}
\end{center}

\caption{Cyclic Proof with Mutually Inductive Predicates}
\label{mutindcylicproof}
\end{minipage}

}
\end{figure*}

\begin{figure*}
\frame{
\begin{minipage}{0.96\textwidth}
\bigskip

\begin{center}
\includegraphics[scale=0.6]{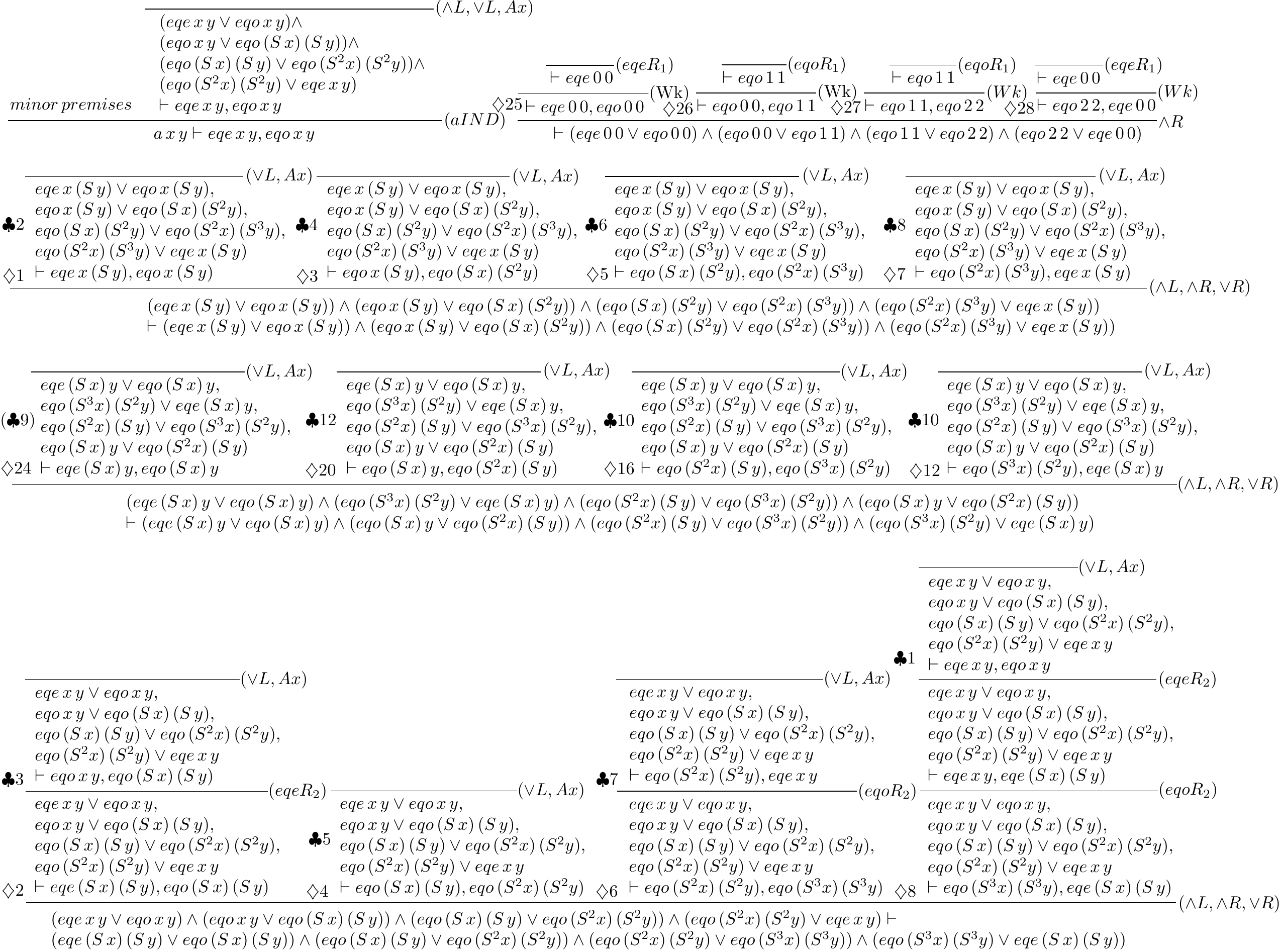}
\end{center}

\caption{Inductive Proof with Mutually Inductive Predicates, part 1}
\label{mutindindproof1}
\end{minipage}

}
\end{figure*}

\begin{figure*}
\frame{
\begin{minipage}{0.96\textwidth}
\bigskip

\begin{center}
\includegraphics[scale=0.6]{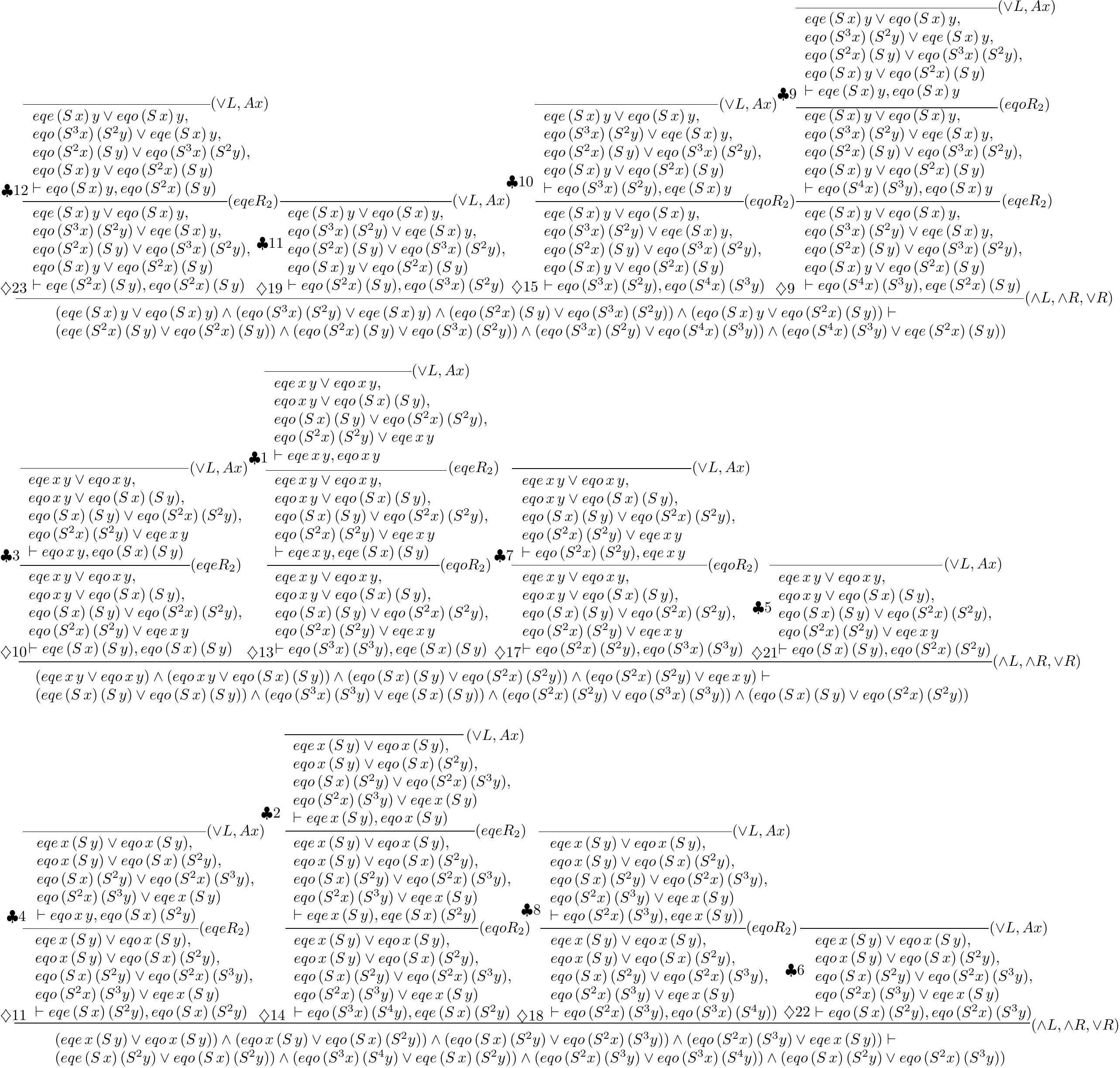}
\end{center}

\caption{Inductive Proof with Mutually Inductive Predicates, part2}
\label{mutindindproof2}
\end{minipage}

}
\end{figure*}

Let ${\cal D}=({\cal S},{\cal R},{\cal P},{\cal C})$,
and let $P_i(\ot_i(\ox))$ be a tagged inductive predicate
such that various renamings $P_i(\ot_i(\oz))$, for some
fresh set of variables $\oz$, appear as principal formulas in
 c-nodes of $\cal S$. Let
$$
\begin{array}{ll}
{\cal K}_i = 
&
\left\{
{\cal E}(\Theta,P_i(\ot_i(\oy)))[\oz/\ox]\left|
\begin{array}{l}
\Theta\in{\cal S}
\mbox{ is} \\ \mbox{a $c$-node with }\\
cs(\Theta)=P_i(\ot_i(\oy))
\end{array}\right.
\right\}
\end{array}
$$
and
$
{\cal H}_i = 
P_i{\bf z_i} \wedge \bigwedge {\cal K}_i
$
where ${\cal H}_i$ is
the {\it induction hypothesis
for $P_i(\ot_i(\ox))$ extracted from $\cal D$}.
Here, the conjunctions appearing at the top level
of ${\cal H}_i(\ox)$ are distinguished connectives,
as are the connectives introduced by the extract notation.

The following lemma defines the simplest case of our translation
scheme and establishes its correctness.

\begin{lemma}\label{lemmasingleton}
Consider a canonical cyclic proof 
${\cal D}=({\cal S},{\cal R},{\cal P},{\cal C})$ with the following
properties: (a) the endsequent is a $c$-node, and (b) 
${\it Fam}({\cal D})$ is a singleton. The following are true.
\begin{itemize}
\item Denote by $P_i(\ot_i)$ the tagged predicate
representing the principal formula of $\cal D$'s
  endsequent. We apply the {\it ($P_i$IND)} inference rule 
to the same endsequent, with ${\cal H}_{i_j}$ as an
induction hypothesis for each predicate 
$P_{i_j}(\ot_{i_j})\in{\cal F}(P_i(\ot_i))$.
We then apply elimination rules
for all the {\it distinguished connectives in the consequents} of all the
resulting minor premises, we obtain an inductive derivation tree
${\cal D}'$
whose frontier has sequents of the
form ${\cal H},\Gamma\vdash\Delta$, where
$\Theta\equiv\Gamma\vdash\Delta$ 
is a $d$-node in $\cal D$,
and $\cal H$ is the induction hypothesis extracted from
$\cal D$ for the tagged predicate that represents the
case-analysis descendant of $cs(\Theta)$ in $\Theta$.
\item Each sequent ${\cal H},\Gamma\vdash\Delta$
on the frontier of ${\cal D}'$ will be fully discharged
by applying the corresponding $dc$-tree rooted at 
$\Gamma\vdash\Delta$ in $\cal D$, followed by elimination
of remaining distinguished connectives.
\end{itemize}

\begin{proof}
Suppose the end sequent is $\Gamma, P_i({\bf u}) \vdash \Delta$.
The major premise is $\Gamma, {\cal H}_i[{\bf u}/{\bf z_j}] \vdash \Delta$.
Denote the root node of {\cal D} by $root$.
We have that ${\cal E}(root, P_i({\bf u}))$ is in ${\cal H}_i$. We apply appropriately many $({\wedge}L)$ rules followed
by $(Wk)$ to get $\Gamma, (\bigwedge \Gamma \rightarrow \bigvee \Delta) \vdash \Delta$, which can be trivially discharged.

For every production rule $\Phi_{j,r}$ whose conclusion is a predicate what is mutually dependent with $P_i$, we have a minor
premise
$$
\begin{array}{l}
\Gamma,Q_1(\ou_1(\oy)),\ldots,Q_k(\ou_k(\oy)),\\
\hspace*{0.3in}
G_{j_1}[\ot_1(\oy)/z_{j_1}],\ldots,G_{j_m}[\ot_m(\oy)/z_{j_m}]\\
\hspace*{1in}\vdash {\cal H}_i[\ot(\oy)/z_i],\Delta
\end{array}
$$
where $\oy$ is a vector of distinct fresh variables. 

let $S = \{k | 1 \leq k \leq m \wedge P_{j_k} \in {\cal F}(P_i) \}$.

we apply $(Wk)$ and $({\wedge}L)$ appropriate times on those $G_{j_x}, x \in S$ to remove the
distinguished connectives and get:
$$
\begin{array}{l}
\Omega, Q_1(\ou_1(\oy)),\ldots,Q_k(\ou_k(\oy)),\\
\hspace*{0.3in}
P_{j_1}[\ot_1(\oy)/z_{j_1}],\ldots,P_{j_m}[\ot_m(\oy)/z_{j_m}]\\
\hspace*{1in}\vdash {\cal H}_i[\ot(\oy)/z_i]
\end{array}
$$
where
$\Omega = (\bigcup_{x \in S} {\cal K}_{j_x}[\ot_x(\oy)/ {\bf z_{j_x}}])$.

We apply $({\wedge}R)$ appropriately many times to break all the distinguished connectives and get
branches of the form:
$$
\begin{array}{l}
\Omega, Q_1(\ou_1(\oy)),\ldots,Q_k(\ou_k(\oy)),\\
\hspace*{0.3in}
P_{j_1}[\ot_1(\oy)/z_{j_1}],\ldots,P_{j_m}[\ot_m(\oy)/z_{j_m}]\\
\hspace*{1in}\vdash F[\ot(\oy)/z_i]
\end{array}
$$ 
where $F \in {\cal K}_i \cup {P_i{\bf z_i}}$.

Suppose $F = P_i{\bf z_i}$, notice that all the premises for the right introduction rule corresponding to $\Phi_{i,r}$ are in the context,
so we can discharge it by applying the right introduction rule, and then the (Ax) rule for all the premises.

Otherwise, $F \in {\cal K}_i$. Therefore we have a $\Theta$ such that ${\cal E}(\Theta,P_i(\ot_i(\oy)))[\oz/\oy]$, and
${\cal S}(\Theta) = \Gamma, P_i(\ot_i(\oy)) \vdash \Delta$. So $F[\ot(\oy)/z_i] =  \bigwedge \Gamma \rightarrow \bigvee \Delta$.

We apply the $({\rightarrow}R)$, $({\wedge}L)$, $({\vee}R)$ appropriate times to remove the distinguished connectives and get
the sequent 
$$
\begin{array}{l}
\Omega, \Gamma[\ot(\oy) / \oy], Q_1(\ou_1(\oy)),\ldots,Q_k(\ou_k(\oy)),\\
\hspace*{0.3in}
P_{j_1}[\ot_1(\oy)/z_{j_1}],\ldots,P_{j_m}[\ot_m(\oy)/z_{j_m}]\\
\hspace*{1in}\vdash \Delta[\ot(\oy) / \oy]
\end{array}
$$
which is denoted by $L$.

Let $vp$ be the premise of $\Theta$ in {\cal D} that corresponds to the production rule $\Phi_{i,r}$, then
$vp$ is a d-node, and
according to the casesplit rule schema, 
${\cal S}(vp)$ is
$$\begin{array}{l}
\Gamma[\ot(\ox)/\oy],Q_1(\ou_1(\ox)),\ldots,Q_k(\ou_k(\ox)),\\
\hspace*{0.5in}P_{j_1}(\ot_1(\ox)),\ldots,P_{j_m}(\ot_m(\ox))
\vdash\Delta[\ot(\ox)/\oy]
\end{array}$$

Let $\Gamma' \vdash \Delta'$ = ${\cal S}(vp)$, then $L[\ox / \oy] = \Omega, \Gamma' \vdash \Delta'$, exactly the
shape we want.

In the process of applying the corresponding dc-tree to the frontier, there can be 3 cases when we reach
the leaf of the dc-tree. If we hit a leaf, it's discharged. If we reach a casesplit, it's necessarily on some
case-analysis descendant of $cs(\Theta)$, say $P_i$. ${\cal H}$ must contain an extracted induction hypothesis for $P_i$.
Therefore we can trivially discharge it. If we reach a cyclic leaf, since the cyclic proof is canonical, it must be identical to
some c-node. By the same argument for the casesplit case, we can discharge it.
\end{proof}

\end{lemma}
\bigskip

\begin{exam}
Consider the following inductive definitions:
$$
\frac{}{a\,0\,0}\quad
\frac{b\,x\,y}{a\,x\,(S\,y)}\quad
\frac{c\,x\,y}{a\,(S\,x)\,y}
\quad
\frac{a\,x\,y}{b\,(S\,x)\,y}\quad
\frac{c\,x\,y}{b\,(S(S\,x))\, y}
$$

$$
\frac{a\,x\,y}{c\,x\,(S\,y)}\quad
\frac{b\,x\,y}{c\,x\,(S(S\,y))}
$$

$$
\frac{}{{\it eqe}\,0\,0}\quad
\frac{eqo\,x\,y}{{\it eqe}\,(S\,x)\,(S\,y)}
\quad
\frac{}{{\it eqo}\,1\,1}\quad
\frac{{\it eqe}\,x\,y}{{\it eqo}\,(S\,x)\,(S\,y)}
$$
and the sequent $a\,x\,y\vdash{\it eqe}\,x\,y,\,{\it eqo}\,x\,y$.
Figure~\ref{mutindcylicproof} presents
a cyclic proof for this sequent, whereas
Figures~\ref{mutindindproof1} and~\ref{mutindindproof2}
present its inductive proof. The inductive proof is obtained 
by simply mechanically applying the translation scheme
described in Lemma~\ref{lemmasingleton}, and is intentionally
complicated so as to showcase the scheme in its generality.
First, we note in Figure~\ref{mutindcylicproof} that
the $c$- and $d$-nodes are marked with $\clubsuit$ symbol
and $\diamondsuit$ symbols. $Dc$-trees are easy to identify, as most
of them have a single branch. For instance, the node marked
with $\diamondsuit$15 is the root of the $d$-tree that applies the
rules {\it (eqoR2)} and {\it (Subst)}.
The cyclic proof contains several shortcuts. Repeating, non-cyclic
nodes are marked with the $\heartsuit$ symbol, in order to save
space. They simply indicate
that the corresponding proof subtree can be pasted at the
corresponding {\it (Idem)} node to obtain a full blown canonical
proof. 

To create an inductive proof, we need to create induction hypotheses
first. For instance, the induction hypothesis for $a\,x\,y$ would be
created from all the $c$-nodes that have a variant of $a\,x\,y$ in
the antecedent. Thus, ${\cal H}_a$ is
$$
\begin{array}{c}
({\it eqe}\,x\,(S\,y)\vee {\it eqo}\,x\,(S\,y))
\wedge\\
({\it eqo}\,x\,(S\,y)\vee{\it eqo}\,(S\,x)\,(S^2\,y))
\wedge\\
({\it eqo}\,(S\,x)\,(S^2\,y)\vee
{\it eqo}\,(S^2\,x)\,(S^3\,y))\wedge\\
({\it eqo}\,(S^2\,x)\,(S^3\,y)\vee{\it eqe}\,x\,(S\,y))
\\\wedge a\,x\,y
\end{array}
$$
Induction hypotheses for $b\,x\,y$ and $c\,x\,y$ can be obtained in a
similar manner, and they appear in the consequents of the minor
premises in the inductive proof.
In the general case, the predicate itself should be part of 
its own induction hypothesis. However, this is not always necessary,
as the predicate instance does not always play a role in the
proof. Since it is the case with the current proof, to simplify the
proof, we have ommitted
the predicates $a\,x\,y$, $b\,x\,y$ and $c\,x\,y$ from their inductive
hypotheses. 

Once the induction hypotheses have been created, the induction
principle can be applied, creating 7 minor premises. Each of the
premises is shown separately. To each premise we can apply rules that
eliminate distinguished connectives that appear in the consequent of
sequents. After this step, all resulting inductive sequents will have a
correspondent in the cyclic proof. The correspondence is indicated by
using the same markings. For instance, node $\diamondsuit$14 in the
cyclic proof corresponds to node $\diamondsuit$14 of the inductive
proofs. The inductive sequent $\diamondsuit$14 looks very similar
to the corresponding cyclic one, but has an extra induction
hypothesis, the one for $c\,x\,y$, in its antecedent. Obviously, the
sequence of rules applied to the cyclic $\diamondsuit$14 can also be
applied to the inductive $\diamondsuit$14, resulting in a node that
corresponds to $\clubsuit$2. This sequent can be discharged by further
elimination of distinguished connectives. It is clear why this
discharge is possible. The consequent of $\clubsuit$2 was used in
creating an induction hypothesis for $b\,x\,y$, and is present in all
antecedents of the subtree rooted at $\diamondsuit$14. This type of
arrangement is used consistently, and will lead to discharging all
minor premises using the $dc$ trees contained in the cyclic tree.
\end{exam}

Given a cyclic proof $\cal D$,
and a family of tagged mutually inductive predicates
$\Psi$, consider 
the result of removing from $\cal D$ all the
$dc$-trees whose root $\Theta$ has the property
that $cs(\Theta)\in{\Psi}$. This set is in
general a set of cyclic derivation trees. Consider a tree $\cal T$ in
this set, and let us examine its frontier. There are two types of
reasons why $\cal T$ is no longer a valid cyclic proof: (a) the
companion of a cyclic leaf now appears in a different tree of the set
(i.e. the cyclic leaf is broken) ;
or (b) a node that used to be the root of a sub-proof has now become a
leaf -- we shall call such nodes {\it open}. Now, we can fix
the broken cyclic leafs by pasting at that node 
a copy of the subtree rooted at the
old companion. Doing this repeatedly will eliminate
the broken cyclic leafs, while possibly increasing the number of open
nodes. We denote by ${\it Erase}({\cal D},{\Psi})$ the set
that contains all the derivation trees obtained by removing
$\cal D$ all the
$dc$-trees whose root $\Theta$ has the property
that $cs(\Theta)\in{\Psi}$, and then repairing the
broken cyclic nodes of the resulting subtree.

Consider now a proof tree ${\cal T}\in
{\it Erase}({\cal D},{\Psi})$, for some
cyclic proof $\cal D$, and some family of mutually inductive
predicates $\Psi$. Denote by ${\cal O}$ the set of open
leafs in $\cal T$, and let $H$ be the formula
$$\bigwedge_{\Gamma\vdash\Delta\in{\cal O}}
(\bigwedge\Gamma\rightarrow\bigvee\Delta),$$
where the connectives introduced in $H$ are distinguished.
Then, the tree ${\cal T}'$ obtained from
$\cal T$ by adding $H$ to the antecedent
of every sequent can be easily converted into
a cyclic proof by elimination rules for the 
distinguished connectives at the formerly open nodes. 
We shall denote the cyclic proof thus obtained 
by ${\it Fix}({\cal T})$.

\begin{lemma}\label{lemmaind}
Consider a canonical cyclic proof 
${\cal D}=({\cal S},{\cal R},{\cal P},{\cal C})$ 
whose endsequent is a $c$-node.
Denote by $P_i(\ot_i)$ the tagged predicate
representing the principal formula of $\cal D$'s
  endsequent. 
 We apply the {\it ($P_i$IND)} inference rule 
to the same endsequent, with ${\cal H}_{i_j}$ as an
induction hypothesis for each predicate 
$P_{i_j}(\ot_{i_j})\in{\cal F}(P_i(\ot_i))$.
We then apply elimination rules
for all the {\it distinguished connectives in the consequents} of all the
resulting minor premises, we obtain an inductive derivation tree
${\cal D}'$. There are two types of nodes
on the frontier of $\cal D$.
\begin{itemize}
\item Sequents of the
form ${\cal H},\Gamma\vdash\Delta$, where
$\Theta\equiv\Gamma\vdash\Delta$ 
is a $d$-node in $\cal D$,
and $\cal H$ is the induction hypothesis extracted from
$\cal D$ for the tagged predicate that represents the
case-analysis descendant of $cs(\Theta)$ in $\Theta$.
Each such sequent will be fully discharged
by applying the corresponding $dc$-tree rooted at 
$\Gamma\vdash\Delta$ in $\cal D$, followed by elimination
of remaining distinguished connectives.
\item Sequents of the
form ${\cal H},\Gamma\vdash\Delta$, where
$\Theta\equiv\Gamma\vdash\Delta$ 
is a $c$-node in $\cal D$,
such that $cs(\Theta)\not\in{\cal F}(P_i)$;
however, $\Theta$ is on the frontier
of a $dc$-tree rooted at some sequent $\Theta'$,
with $P_j(\ot_j)\equiv cs(\Theta')\in{\cal F}(P_i(\ot_i))$
and $\cal H$ being the induction hypothesis
used in the application of {\it ($P_i$IND)} for
the case-analysis descendant of $P_j(\ot_j)$
residing in $\Theta'$. Moreover
${\cal H},\Gamma\vdash\Delta$
is the endsequent 
of the cyclic proof
${\it Fix}({\cal T})$
where $\cal T$ is the derivation tree
rooted at $\Theta$ in 
${\it Erase}({\cal D},{\cal F}(P_i(\ot_i)))$.
\end{itemize}

\begin{proof}
For d-node frontiers, discharge them by methods described in Lemma 1.
For c-node frontier case.
We must have collected $\Delta[{\bf z_i}/{\bf u}]$ at some c-node $\Theta'$ in
{\cal D}, where $cs(\Theta') \in {\cal F}(P_i(\ot_i))$, and $\Theta'$ has a premise $vp$,
where ${\cal S}(vp) = \Gamma \vdash \Delta$. Now $\Theta$ is on the frontier of
a dc-tree rooted at $\Theta'$, and ${\cal H}$ is the induction hypothesis used in the application
of $(P_iIND)$ for the case-analysis descendant of $P_j(\ot_j)$ residing in $\Theta'$. In fact,
$\Theta$ is a premise of $\Theta'$. In $Erase({\cal D}, {\cal F}(P_i(\ot_i)))$, $\Theta$ is not
a descendant of any other node. Therefore it is the endsequent of the cyclic proof $Fix({\cal T})$, where ${\cal T}$ is
the derivation tree rooted at $\Theta$.

\end{proof}
\end{lemma}

It is worthy of noting at this point that the presence
of the predicate $P(\ot)$ in its own inductive hypothesis is indeed
necessary, so as to ensure that for each $d$-node, its corresponding
inductive node is a strict weakening. This furthermore ensures
that any sequence of rules applicable to the $d$ node is also
applicable to its corresponding inductive node. For instance,
let us examine closely the cyclic proof in
Figure~\ref{tpluscyclicproof}, and its inductive
translation given in Figure~\ref{tplusinductiveproof}.
In this proof, the inductive hypothesis for the predicate {\it nat}
is ${\it nat}\,k\wedge{\it tplus}\,k\,0\,k$. Had we not added ${\it
  nat}\,k$ to the inductive hypothesis, the weakening just above the
right premise of the cut rule would not have been possible. That is to
say that the $d$-tree rooted at the $\diamondsuit$2 node in the cyclic
proof would not have been applicable to the corresponding node in the
inductive proof, invalidating both Lemmas~\ref{lemmasingleton}
and~\ref{lemmaind}. \bigskip

\begin{theo}[Former Conjecture 7.7]
Let $\cal D$ be a canonical
CLKID$^\omega$ proof with endsequent $\Gamma\vdash\Delta$.
Then, there exists an LKID proof of $\Gamma\vdash\Delta$. 
\end{theo}

\begin{proof}
First, we isolate and remove from $\cal D$
the subtrees rooted at the shallowest c-nodes (the ones that do not
have other c-node ancestors). The result of the removal is a
derivation tree that is valid in LKID. If we convert the cyclic c-node
rooted proof trees that were removed into inductive proof trees, and
then paste them back, we obtain a valid inductive proof for the
original sequent. Thus, we can focus only on converting 
c-node rooted proof trees into inductive ones. 
Following Lemma~\ref{lemmaind}, we apply
the ($P_i${\it IND}) induction rule to the endsequent
$\Theta\equiv\Gamma\vdash\Delta$, where
$P_i = cs(\Theta)$ in $\cal D$. After
elimination of distinguished connectives, we obtain
a valid inductive derivation tree ${\cal D}'$. On the
frontier of ${\cal D}'$ we have sequents of the form
${\cal H},\Gamma'\vdash \Delta'$ which can be in 
either of the following two situations.
The first situation is that $\Gamma'\vdash\Delta'$
is the root of a $d$-tree $\cal T$ in $\cal D$, 
in which $\cal T$ is applicable to
 ${\cal H},\Gamma'\vdash \Delta'$, and leads sequents that can be
trivially discharged by application of elimination rules for
distinguished connectives. Thus, part of the outstanding sequents
after the application of the inductive principle will be discharged.
The second situation is that $\Gamma'\vdash\Delta'$
is a $c$-node in $\cal D$. In this case,
${\cal H},\Gamma'\vdash \Delta'$ cannot be immediately discharged.
However, each of the sequents that cannot be immediately discharged
is in the set of endsequents of proofs in the set
$\{{\it Fix}({\cal T})|
{\cal T}\in{\it Erase}({\cal D},P_i(\ot_i))\}$.
Now, each of these proofs is strictly smaller than $\cal D$, and can
be recursively converted into an inductive proof by 
Lemma~\ref{lemmaind}. Each of the
inductive proofs can be pasted at the right place in the current
inductive derivation tree, to produce a valid inductive proof tree for
the original endsequent, which proves the theorem. Most notably, our
(meta)proof is in fact a cyclic proof!
\end{proof}

\section{Further Work}

In \cite{BrotherstonJLC}, the discussion before Conjecture~7.7 states
that cut is probably not eliminable from CLKID$^\omega$. This
statement also appears as Conjecture~5.2.4
in \cite{thesis/brotherston}. We believe that this is true, since the
sequent $nat\,k\vdash{\it tplus}\,k\,0\,k$, which has a non-cut free
proof, as shown in
one of our examples, also has a unique and infinite derivation in 
a cut-free CLKID. Both uniqueness and non-finiteness are 
easy to establish, and show that, since there exists a sequent
with no cut-free proof, but provable in the presence of cut, 
the cut inference rule is not admissible in CLKID$^\omega$.

Moreover, we belive that it is possible to define a notion of
anchored cut similar to Gentzen's (cf. \cite{Buss98}). Indeed, the
only type of cuts that really are needed are the ones where the
cut formulas are inductive
predicates that have previously been subjected to case-analysis rules;
all other cuts can be easily eliminated.

Thus, by
defining free cuts as non-anchored cuts, a
free-cut elimination result can be obtained, leading to a natural analogue
of the subformula property. It would then naturally follow that, due
to the translation scheme presented in this work, the induction
calculus LKID would enjoy a form of subformula property too. This is
contrary to popular belief (cf. \cite{conf/sls/martinlof71}), but not
necessarily surprising at this point, since it has become clear that
the induction hypotheses, previously believed to be completely
arbitrary, are made up exclusively from sequents appearing in the
cyclic proof, and can, in the presence of a free-cut elimination
result, be obtained by recombining parts of the endsequent.

\bibliography{lics2011}
\bibliographystyle{plain}

\end{document}